# The Vapor-Solid-Solid Growth of Ge Nanowires on Ge (110) by Molecular Beam Epitaxy

Zhongyunshen Zhu,[‡,1,2] Yuxin Song,[‡,1*] Zhenpu Zhang,[1,2] Hao Sun,[1] Yi Han,[1,3] Yaoyao Li,[1] Liyao Zhang,[1] Zhongying Xue,[1] Zengfeng Di,[1] and Shumin Wang[1,2,4*]

[1]State Key Laboratory of Functional Materials for Informatics, Shanghai Institute of Microsystem and Information Technology, Chinese Academy of Sciences, Shanghai 200050, China
[2]School of Physical Science and Technology, ShanghaiTech University, Shanghai 201210, China
[3]University of Chinese Academy of Sciences, Beijing 100190, China
[4]Department of Microtechnology and Nanoscience, Chalmers University of Technology, Gothenburg 41296, Sweden
[‡]These authors contributed equally to this work.
*Corresponding authors: songyuxin@mail.sim.ac.cn, shumin@mail.sim.ac.cn

**Abstract**
We demonstrate Au-assisted vapor-solid-solid (VSS) growth of Ge nanowires (NWs) by molecular beam epitaxy (MBE) at 220 °C, which is compatible with the temperature window for Si-based integrated circuit. Low temperature grown Ge NWs hold a smaller size, similar uniformity and better fit with Au tips in diameter, in contrast to Ge NWs grown at around or above the eutectic temperature of Au-Ge alloy in the vapor-liquid-solid (VLS) growth. Three growth orientations were observed on Ge (110) by the VSS growth at 220 °C, differing from only one growth direction of Ge NWs by the VLS growth at a high temperature. The evolution of NWs dimension and morphology from the VLS growth to the VSS growth is qualitatively explained via analyzing the mechanism of the two growth modes.

**Introduction**
Group-IV nanowires (NWs), including silicon (Si) NWs, germanium (Ge) NWs and germanium-tin (GeSn) NWs, meet the demand of current Si-based industry for both electronics[1-3] and photonics[4,5]. Especially, compared with Si NWs, Ge NWs have promising electronic properties[6] and play an increasingly important role in various fields, such as high speed transistors,[7] photodetectors[8] and the anode material of high capacity lithium-ion batteries.[9] Conventional Ge NW growth with gold (Au)-assisted vapor-liquid-solid (VLS) process takes place above the Au-Ge eutectic temperature ($T_e \approx 361$ °C).[10-12] However, for the VLS growth of NWs, it is hard to control the NW size and position due to the surface migration and coalescence of the catalysts during the growth. Furthermore, NW tapering or kinking is more likely to occur in the VLS growth to affect the shape of NWs.[13] These characters of VLS opposes to the requirements for the practical application of the bottom-up NWs in large scale integrate circuits, which requires selective growth of NWs in designed positions and sizes. From this point of view, the vapor-solid-solid (VSS) growth of NWs can be attractively worth for taking into practice to overcome the drawback of VLS, since both the size and position of the solid catalysts can be determined expectedly thanks to the solid phase of catalysts during the growth, minimizing the catalyst migration and coalescence.[14, 15] Moreover, the VSS process of Au-Ge system can significantly lower the growth temperature, raising the compatibility with Si-based microelectronics. Meanwhile, for group-IV photonics, low growth temperature of Ge NWs can promote the possibility to incorporate Sn, while GeSn NW is potential to become a direct bandgap material. Thus, the VSS growth of Ge NWs is of importance for investigation. The VSS growth of Ge NWs was realized with Au catalyst at a temperature lower than $T_e$ under low growth rate by utilizing ultra-high vacuum (UHV) chemical vapor deposition (CVD)[16] or physical vapor deposition (PVD)[17], showing the same growth direction as the VLS process on Si (111) or Ge (100) substrate, respectively.

Here, we demonstrate Au catalyzed Ge NWs grown on Ge (110) substrates through the VSS process by molecular beam epitaxy (MBE). The growth temperature used is lowered down to 220 °C, which is much lower than $T_e$ without the impact of the Gibbs-Thomason effect[18] and satisfies the compatibility with the temperature window for Si-based integrated circuit. For comparison, we also grow Ge NWs at above $T_e$ as reference. Our result shows that the diameter as well as the density of Ge NWs decrease with reducing growth temperature from above $T_e$ to below $T_e$, whereas the size uniformity is not changed. Also, the VSS process leads to NWs in three <110> growth directions simultaneously: vertical, lateral and 30°-tilted to surface, in contrast to only one vertical direction in the VLS growth. Besides, the diameter of the Au particles approximately fits that of the NWs in the VSS process, but becomes much smaller during the VLS growth. All these morphological differences in both the growth orientation and the NW diameter can be explained by the fundamental growth mechanism in the VSS and the VLS processes. The Ge (110) substrate is selected because Ge NWs grow along <110> by VLS without atomic hydrogen protection,[19] and most of the Ge NWs are vertical to the Ge (110) substrate.[11]

**Experiment**

Ge NWs were grown by a DCA P450 solid source MBE system. Prior to the growth, Ge substrates were degassed at 400 °C for 30 minutes and further heated to 650 °C for an hour to remove native oxides. Then, 2 nm thick Au film was deposited onto cleaned Ge substrates at room temperature by electron beam evaporation and annealed at 450 °C for 30 minutes to form nanoparticles. Finally, Ge NWs were grown on Ge (110) substrates under different growth temperatures but a fixed Ge growth rate of ~0.010 nm/s and total growth time of 3.5 hours under UHV of ~$1.0 \times 10^{-9}$ Torr. At this low growth rate, the VSS process is responsible for the Ge NW growth at the temperature far below $T_e$. Because of the distinction in growth mechanism for Ge NWs grown above and below $T_e$, we choose growth temperatures of 220 °C, 290 °C, 380 °C and 470 °C to exhibit morphologic transition. As described before, a traditional growth temperature of Au catalyzed Ge NWs is above $T_e$ and ranges from 370 °C to 510 °C[12] with the VLS growth mode, indicating that Au catalyst forms as liquid droplets. However, when the growth temperature decreases to much below $T_e$, the mechanism at a low growth rate is the VSS mode, leading to formation of solid Au nanoparticles. Moreover, two-step growth from 470 °C to 220 °C (growing at 470 °C for 1.5 hours then cooling down to 220 °C to grow for 2 hours) is experimented to study the transition from VLS to VSS process. Post-growth characterization for morphology of Ge NWs employs scanning electron microscope (SEM) at different angles. Scanning transmission electron microscope (STEM) and energy dispersive spectrum (EDS) mapping were used to show distribution of Au and Ge in Ge NWs.

**Results and Discussion**

Ge NWs on Ge (110) substrates were successfully grown at 220 °C by VSS MBE with Au catalyst. Figures 1(a) and (b) reveal the SEM images observed from top view and 35°-tilted to the sample, showing a low density and small size in diameter of NWs, respectively. Figures 1(c)-(e) show the STEM picture of Ge NWs and EDS mapping of (d) Au and (e) Ge, respectively. The Au tip and the Ge NW are easily distinguished. Another large Au particle with a diameter lager than 50 nm is observed in the position close to the Ge NW. The high background of Au found in the Ge NW and substrate is probably related to the contamination by Au during focused ion beam preparation of the STEM specie, since Au is easily sputtered and coated on sidewalls of the specie. On the other hand, there is little amount of Ge in the Au tip, demonstrating that it is hard for Ge to diffuse into Au tip during the VSS growth.

For comparison, we have investigated Ge NW growth at the selected temperatures on Ge (110) substrate with histogram on size, average NW diameter as well as SEM images shown in Fig. 2. Figure 2(a) shows the histogram of Ge NWs. The NWs grown at 220 °C reveal a symmetric and narrow size distribution centered at 40-60 nm in diameter, while the NWs grown at 380 °C have an asymmetric and broad size distribution with the weight on 100-120 nm in diameter. Furthermore, the average diameter values of not only the Ge NWs but also the catalyst droplets are calculated by counting over 30 NWs per sample and shown in Fig. 2(b). The average diameter of Ge NWs increases remarkably with growth temperature from 290 °C to 380 °C but otherwise changes slightly. The diameter values of Ge NWs during the VLS process are about twice than that during the VSS process. The corresponding standard deviation also tends to increase with increasing growth temperatures above $T_e$, but remains nearly a constant if normalized by the mean values. It indicates that the uniformity of Ge NWs is approximately invariant at each growth temperature. Additionally, the similar trend of the diameters of catalyst droplets with varying the growth temperature is seen in Fig. 2(b). Interestingly, the catalysts have roughly same diameters as NWs below $T_e$, while the diameters of droplets are much smaller than those of NWs above $T_e$. This statistical data well fits our assumption that the droplets stay only on one of the top {111} facets in the VLS process, as the schematics shown in the insets of Fig. 2(b). The droplet diameter $d$ and the NW diameter $D_{VLS}$ have the relation of $D_{VLS} \approx 2d \sin(109.5°/2)$.

Figure 2(c) is SEM images including a single Ge NW grown at designed growth temperatures on Ge (110). Both the NW diameter and length increase when the growth mode transits form VSS to VLS in consistent with the result in Fig. 2(b). This dramatic change in volume is primarily determined by the temperature dependent diffusion length of Ge atoms during MBE NW growth.[12, 20] It was reported that ~430 °C is the optimal temperature for Ge NW growth by VLS.[11, 12] Thus the Ge diffusion length is supposed to reach maximum at this growth temperature. A higher temperature may restrain Ge diffusing towards Au droplets, but assist Ge to grow on the area between NWs. From this point of view, we can explain that the size of Ge NWs becomes similar at the growth temperatures of 380 °C and 470 °C in Fig. 2(c), whereas the density of Ge NWs grown at 470 °C is lower than the situation of 380 °C. The diffusion length is further limited by the ultra-low growth temperature far below $T_e$. As a result, less amount of Ge is capable of travelling from the impingent plane or sidewalls of NWs to the interface between Au catalyst and Ge to form NWs, causing a small size of NWs. In addition, the interface of the solid Au nanoparticle and the NW seems to be tilted at the growth temperature of 290 °C. For the sample grown at 220 °C, we can see the Au nanoparticle is roughly round, which implies the same case as the 290 °C grown sample.

Further, we compare growth orientation of Ge NWs at temperatures below and above $T_e$, shown in Fig. 3(a) and (b), respectively. It was experimentally proved that Ge NWs always grow along <110> directions with a low growth rate.[11, 21] Three dominant orientations of Ge NW growth are observed, delineated in Fig. 3(a) (we assume that [110] is perpendicular to the Ge (110) substrate). The solid arrow lines denote axes on Ge (110) while the dash arrow line is [10-1]. At the growth temperatures above $T_e$, most of Ge NWs prefer to grow along the [110] and other four equivalent directions: [011], [01-1], [101] and [10-1], are rarely observed, in agreement with the result of *Schmidtbauer et al.*[11] Interestingly, at a low growth temperature, lateral NWs can appear along the [1-10] direction. Scrutinizing through different positions of the NW sample grown at 220 °C, we find that the NWs with 30° inclination possess long length, thin diameter and small Au tips in common, both shown in Fig. 1(a) and Fig. 3(a). This result probably indicates that the small Au nanoparticle will induce the nucleation occurring along [10-1] (other three equivalent directions are [-101], [1-10] and [-110]), whose direction is 30°-tilted to the substrate. Besides, compared to the vertical ones, the additional way to form the inclined NWs is that some of Ge atoms can reach the sidewall of NWs without diffusing from the substrate, leading to more efficient amount of Ge enter the NWs. In Fig. 3(b), some of NWs have a droplet on the their top or sidewall, but some NWs have lost it possibly because of the catalyst spreading[19] and Ostwald ripening process[22] during the VLS growth.

To investigate the transition from the VLS growth to the VSS growth, a sample with two-step growth has been prepared. We first grew Ge NWs on Ge (110) at 470 °C for 1.5 hours, followed by the growth at 220 °C (cooling down to 220 °C directly) for 2 hours, keeping the same total growth time as for the previous samples. The 45°-tilted SEM pictures illustrated in Fig. 4 show the abrupt boundaries of the two steps marked by white dash lines, regardless of the orientation. The white arrow lines denote the NW growth direction under the VSS process. As expected, the second step of growth at 220 °C displays identical three main growth directions, which are [110], [10-1] and [1-10] in Fig. 4(a), (b) and (c), respectively, like the result in Fig. 3(a). Moreover, almost every NW grows along the [110] in the first step due to the VLS growth, the same as the result of the NW growth at 470 °C. This match of orientation between the two-step growth and the direct growth at 220 °C or 470 °C suggests that the VSS process occurs when Ge NWs are grown at 220 °C and the VLS process occurs at 470 °C. It is notable that the NW diameters in the first step approximately double those in the second step as shown in Fig. 4, being consistent with the statistical consequence in Fig. 2(b).

From the standpoint of elementary processes in the NW growth model by *Tersoff*,[23] the stability and the size of droplets have a significant impact on the formation of NWs in the VLS process. To simplify the explanation, we use a 2D model shown in Fig. 5(a).[23] The system is stable if the contact angle ($\theta_c$) of the droplet situates inside the range marked by two dash lines. Additionally, sidewalls of the NW are {111} facets which have the minimum surface energy. Initially, an Au droplet lies on the surface to catalyze growing a Ge pedestal exposed by {111} facets on the Ge (110) substrate illustrated in Fig. 5(b). As the system evolves, the interface becomes so large that $\theta_c$ decreases outside the stable range to form a step marked by the blue circle of Fig. 5(c) and then the interface will tend to diminish to a small one and the $\theta_c$ increases, as shown in Fig. 5(c). When the $\theta_c$ falls outside the stable range again, the droplet will depin from the right trijunction, indicated by the arrow in Fig. 5(c). The reason that vertical NWs dominate while their inclined counterparts are rare, i.e. the depinning process does not cross the left trijunction, is unclear but may originate from the low probability. During further NW growth, the depinned droplet slides along the (11-1) facet until forming a new step shown in Fig. 5(d). Referring to the dynamics analysis of the VLS growth,[24] when the overall growth rate is low which is the case in this work, the larger wetting facet has a higher growth velocity than that of the smaller one. This will eventually lead to a symmetric growth front shown in Fig. 5(e). As stated previously, in addition to Fig. 2(b), we assume that most of Au droplets finally solidify to stay on only one of the two top {111} facets at high growth temperatures. This phenomenon may be explained by the jumping-catalyst dynamics of the alternating growth from one facet to another due to the mismatch between the catalyst size and the NW diameter.[25] This mismatch can be explained as follow. The droplets experience Au atom evaporation and Ge atom out-separation (for keeping the saturated Ge in Au) due to the UHV character of MBE growth, in addition to out-diffusion towards the Ge NW underneath to combine with a big droplet owing to Ostwald ripening or dissolved by the NW sidewall.[19] The long growth time of several hours can cause the gradually decreasing size of the droplet till an unstable state because of small $\theta_c$, shown in Fig. 5(f). Then the Au droplet jumps to one of the two top facets to reach a new stable state as sketched in Fig. 5(g). Continuing growth will lead to the Au droplet hopping over the two top facets alternately as demonstrated by *Tersoff et al.* in the movie M5 in the supporting material of Ref. 25. The growing process of the inclined NWs is the same as the above mechanism but rarely occurs in the VLS growth.

Differently, in the VSS growth, the phase of Au nanoparticle is solid which has an abrupt interface with the NW tip. In this case, we assume that a catalytic nanoparticle may only sit on one {111} facet of a pedestal, as shown in Fig. 5(h) with the tilted catalyst/NW interface satisfying with the result in Fig. 2(c). Because of an abrupt solid/solid interface throughout the growth, the nanoparticle may tightly attach to the {111} facet of the pedestal to advance one of five equivalent <110> growth directions, including a vertical one, two 30°-tilted directions and two lateral directions. Figure 5(i) just exhibits one situation as an example. Thus, the process during two-step growth has identical growth directions in Fig. 3(a). As addressed previously, in Fig. 2 and Fig. 4, the diameter of a NW equals that of the solid Au nanoparticle in the VSS growth, but is larger than that of the Au catalyst in the VLS growth.

In accordance with the mechanism, the VSS growth has a significant impact on the NW size by efficiently decreasing the NW diameter and increasing its uniformity. Both advantages of the VSS growth on NWs morphology can result in a small diameter to enhance the radial quantum confinement. In addition, during the VLS process, because Ge enters Au droplets at high temperatures, the catalyst becomes Ge-supersaturated Au-Ge droplets. Contrarily, with the solid catalyst, the solubility of Ge in Au is lower than that in the VLS process. Solidified Au nanoparticles inhibit foreign Ge atoms injecting into themselves, ensuring the chemical purity of NWs.[14] This property determines that the average diameter of the catalysts in the VLS process is larger than that in the VSS process, as shown in Fig. 2(b). Moreover, holding the point of two-step growth, our result with abrupt interface in the VSS growth is positive for defect-free axial NW heteroepitaxy, such as Si/Ge or GeSn/Ge system since it has been predicted that axial heterostructural NWs can relax the strain significantly without dislocations if the diameter of the top epi-layer is smaller than a critical value.[26]

**Conclusion**

In summary, we have successfully grown Ge NWs on Ge (110) at 220 °C by VSS MBE, and found that the morphology both in growth direction and the dimension are different from these of the VLS process. Compared to the NWs under the VLS process at higher temperatures than $T_e$, both the diameter and length decrease significantly in the VSS process. The diameters of Ge NWs under the VSS process strictly follow the size of Au nanoparticles, leading to a controllable growth. Simultaneously, pure NWs can be obtained from the VSS growth due to the prohibited mixing of Ge with the solid Au catalyst. The two-step growth of Ge NWs from the VLS growth switching to the VSS growth has been realized. Ge NWs grown under the second VSS process show the similarity in three growth orientations and the same diameter as that of the NWs directly grown at 220 °C.

**Acknowledge**

This study was supported by the Natural Science Foundation of China (61404153), the Shanghai Pujiang Program (14PJ1410600), the Key Research Program of the Chinese Academy of Sciences (KGZD-EW-804), the Creative Research Group Project of Natural Science Foundation of China (61321492).

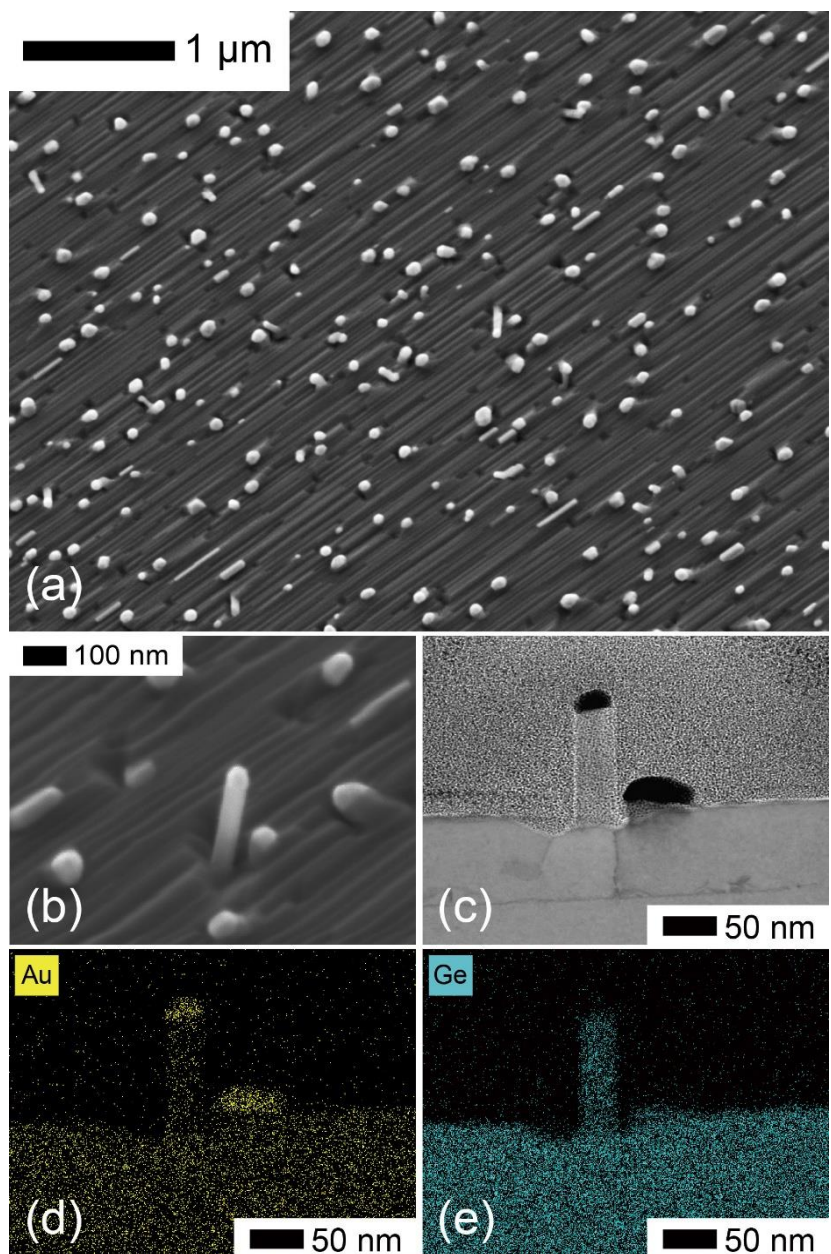

Fig. 1 SEM images of Ge NWs grown at 220 °C on Ge (110) observed from different angles: (a) in-plane and (b) tilted by 35°. (c) STEM image of the Ge NW grown at 220 °C on Ge (110) and the EDS mapping of (d) Au and (e) Ge.

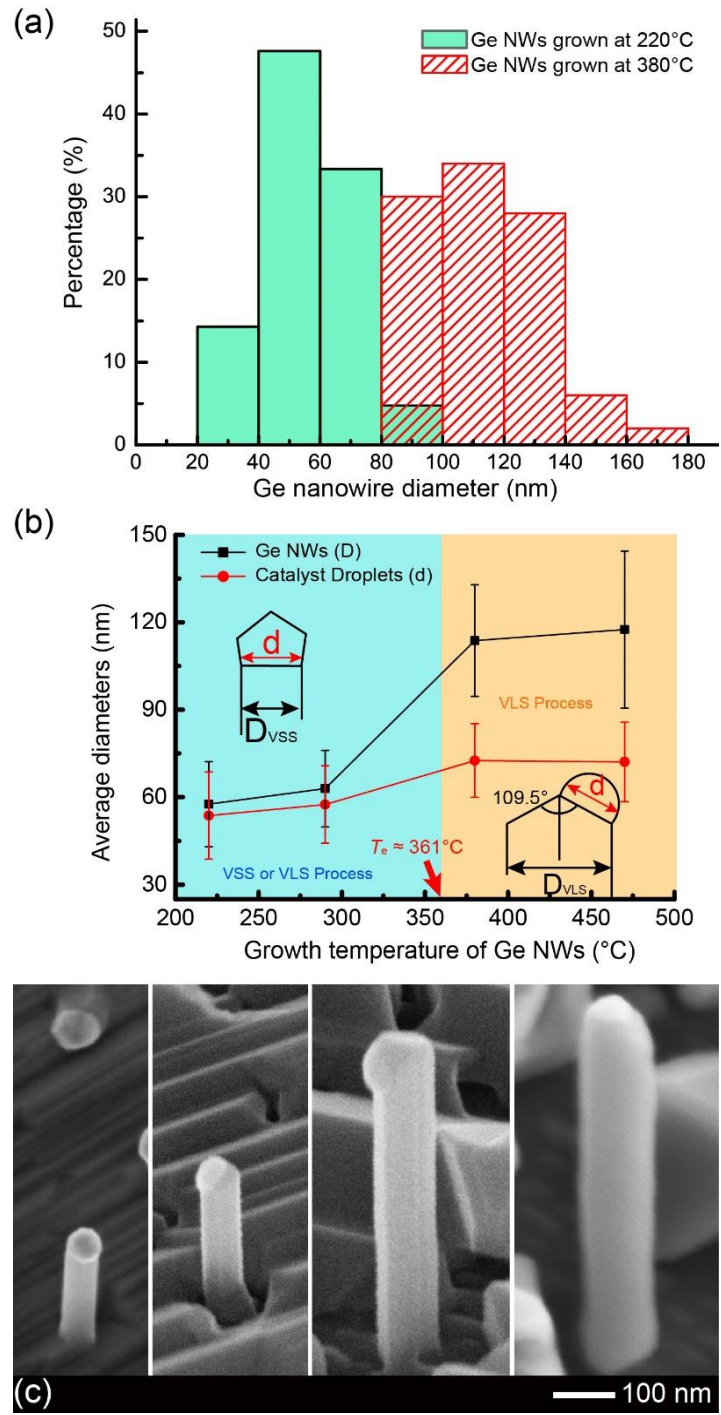

Fig. 2 (a) The histogram of Ge NWs grown at 220 °C (green bars) and at 470 °C (red pattern bars); (b) Average diameter and its error bars of Ge NWs (black squares) and the corresponding Au tips (red dots) as a function of growth temperature. (c) Tilted SEM images of single Ge NW grown at different temperatures on Ge (110).

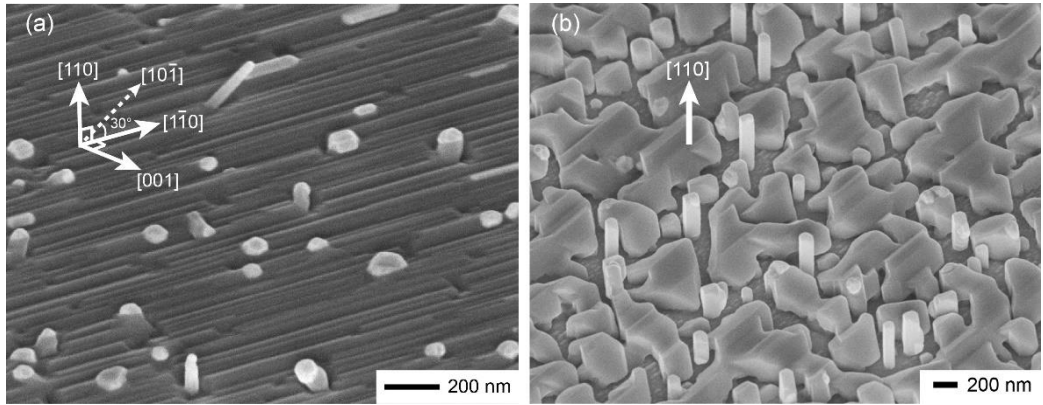

Fig. 3 45°-tilted SEM analysis of Ge NW growth orientation on Ge (110) at (a) 220 °C and (b) 470 °C.

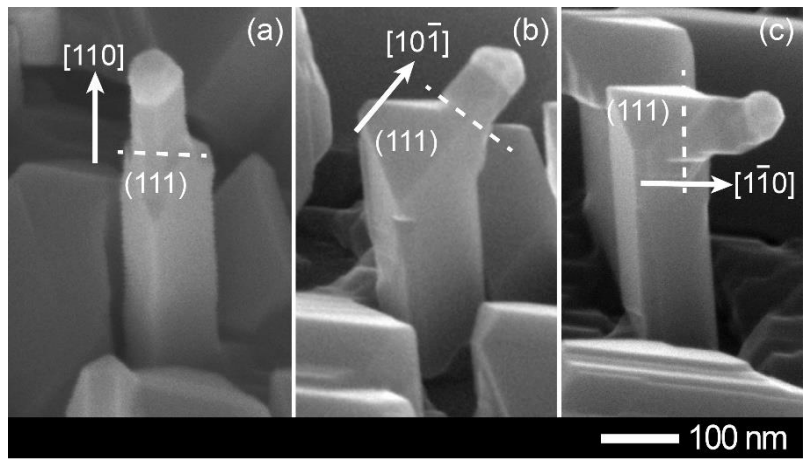

Fig. 4 SEM images of Ge NWs with the two-step growth on Ge (110). The second growth mode is separated from the first one by dash lines and shows three orientations: (a) vertical, (b) 30°-tilted to the surface and (c) in-plane.

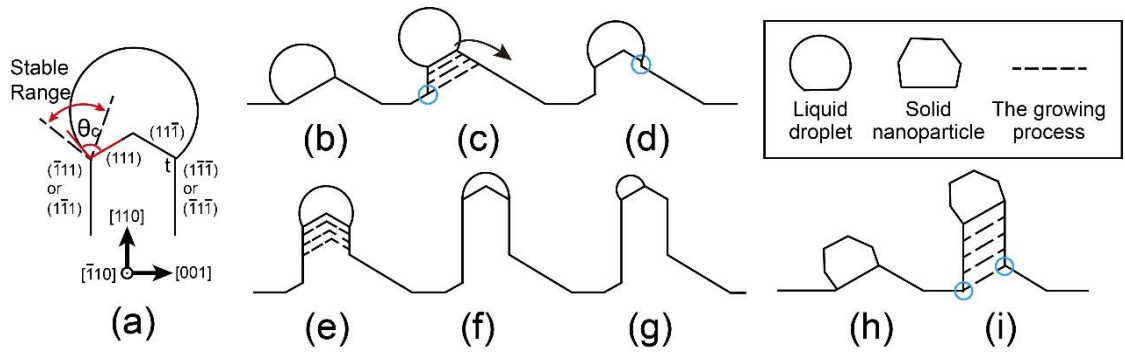

Fig. 5 Illustration of growth mechanism of the VLS and the VSS growth. (a) 2D model from the view illustrated as the orthotropic coordinate axis. The interface contains (111) and (11-1) facets. The sidewalls of the NWs denote any of the {111} facets other than the (111) and (11-1) facets. The trijunction where vapor, liquid, and solid meet is labeled *t*. The contact angle $\theta_c$ is marked by two red lines, while dash lines represent the stable range of $\theta_c$. (b)-(i) are the specific process in the VLS growth (b)-(g) and in the VSS growth (h)-(i) of Ge NWs on Ge (110) substrate. The blue circles denote step forming in the process. We only choose the vertically grown NWs for example.